\documentclass[12pt,fleqn]{article}
\usepackage[margin=1in]{geometry}
\usepackage[fleqn]{amsmath}
\usepackage{mathtools}
\usepackage{amsthm}
\usepackage{amssymb}
\usepackage[sort,comma]{natbib}
\usepackage{graphicx}
\usepackage{placeins}
\usepackage[colorlinks]{hyperref}
\usepackage{setspace}
\usepackage{authblk}
\usepackage{subfigure}
\usepackage{algpseudocode,algorithm,algorithmicx}

\rmfamily
\normalsize
\newcommand*{\nl}[1]{%
  \begin{NoHyper}#1\end{NoHyper}%
}
\hypersetup{
    colorlinks=true,
    linkcolor=blue,
    citecolor=blue
}

\newcommand{\nml}{\noindent} 						
\newcommand{\ml}{\noindent \hspace{1in}}		
\newcommand{\ems}{\vspace{0.005cm}} 				

\newcommand{\bm}{\mathbf}

\newcommand{\prob}{\mathbb{P}}
\newcommand{\skel}{\mathcal{S}}

\graphicspath{{simulations/}}

\author{Samuel Baltz\thanks{\protect\linespread{1}\normalsize Massachusetts Institute of Technology, sbaltz@umich.edu. This paper benefitted from discussions, including several technical improvements and corrections, from Luka Buli{\'c} Bra\v{c}ulj, Edward Foley, Joelle Gross, Christa Hawthorne, Walter R. Mebane, Jr., Scott E. Page, Charles Stewart III, and Gillian Thompson. Fact-checking was performed on computers available through a fellowship from the Michigan Institute for Computational Discovery and Engineering.}}
\title{The probability of casting a pivotal vote in an Instant Runoff Voting election}
\date{\today \\ Word count: 6,930}

\begin{document}
\doublespacing

\maketitle

\singlespacing
\begin{abstract}
	I derive the probability that a vote cast in an Instant Runoff Voting election will change the election winner. I phrase that probability in terms of the candidates' expected vote totals, and then I estimate its magnitude for different distributions of voter preferences. The result is very similar to the probability of casting a pivotal vote in a Single-Member District Plurality election, which suggests that Instant Runoff Voting does not actually increase or decrease voters' incentives to vote strategically. The derivation uncovers a counter-intuitive phenomenon that I call ``indirect pivotality'', in which a voter can cause one candidate to win by ranking some other candidate on their ballot.
\end{abstract}
\doublespacing

\newpage

\section{Introduction}
After a wave of recent reforms, Instant Runoff Voting (IRV) is suddenly a widely used electoral system in major elections, but little is known about how voters behave in this ruleset. Because IRV is a ranked-choice electoral system, some scholars and advocates reason that it allows voters to report their sincere preference order on their ballot \nl{\citep[ch. 5]{gehl20}}. Another possibility, though, is that the ranked options in IRV provide voters with distinct reasons to act strategically, and different tools for doing so \nl{\citep{santucci21,eggers21b}}. Strategic voters attempt to influence election results by supporting more popular candidates, so whether voters have more or less opportunity to influence election outcomes in IRV depends on the probability that a particular ballot changes the result of an IRV election. However, that probability has only been derived in IRV contests of up to four candidates \nl{\citep{bouton13,eggers21b}}. So, how big are voters' actual incentives to vote strategically in IRV?

The core component of strategic voting models is the expected utility of each vote choice, but calculating the expected utility of a ballot in IRV requires knowing the probability that a vote cast in an IRV election will change the election winner. This question of pivotal probabilities was first posed about the more traditional Single-Member District Plurality (SMDP) electoral system more than half a century ago, and various frameworks for estimating the quantity have been explored in detail since then \nl{\citep{riker68,cox94,mebane19,eggers20,vasselai21b}}. Pivotal probabilities have also been extended to proportional systems \nl{\citep{cox96}}, and even the alternative ranked-choice system of Borda count \nl{\citep{baltz22}}. Researchers have even identified the set of all pivotal outcomes in a three- or four-candidate IRV race, and modeled the probability of each outcome arising in a realistic election \nl{\citep{bouton13,eggers21b}}. However, the general question of pivotal opportunities in IRV remains open: it is not known whether or not voters have more or less opportunity to cast pivotal votes in IRV compared to more conventional systems like SMDP.

I derive the probability that a vote cast in a single-winner IRV election changes the outcome of that election, when any number of candidates contest the election, and voters can rank any number of them on their ballots. This is not a straightforward generalization of the 3- or 4-candidate case; those situations were solved by exhaustive enumeration, but the case with any number of candidates requires deriving an expression for the probability of a pivotal event (any situation in which a single ballot could change the election result) as a function of the ballots that are expected to be cast. The resulting equation is exceptionally complicated, which I argue reflects real complications in the electoral system. The derivation also uncovers a property of IRV that I call ``indirect pivotality'', in which a voter can cause one candidate to win by ranking some other candidate, and I argue that this represents a substantive motivation to keep ballot length short in IRV.\footnote{By ``ballot length'' I mean the number of candidates that each voter is allowed to rank to determine the single winner of an IRV election. I will also use ``ballot'' to mean just the section of the ballot relating to a specific IRV election (recognizing that a ballot may have multiple offices on it, but lacking a simple term for ``just the part of a ballot that corresponds to the IRV race under consideration'').} I then show that, under one well-studied method for estimating numerically specific pivotal probabilities, the opportunity to cast a pivotal vote is similar in IRV and SMDP, which suggests that neither system fundamentally encourages strategic voting more than the other.

Of course, the literal expected utility of each vote choice may or may not influence real voting behaviour; exactly how strategic voters are in IRV, and what factors influence their vote choice, is an open question that has been the subject of substantial recent research \nl{\citep{atsusaka23,reilly21,buisseret22,simmons22}}. However, the pivotal probability of an IRV ballot is an important quantity even apart from its potential impact on voting behaviour. The foremost reason is that this probability shapes real outcomes in IRV. If ties tend to be more common in IRV, then voters may find themselves acting pivotally more often, whether they plan to or not (this is naturally a larger concern in smaller electorates). Another reason is that models of voting behaviour often require knowing pivotal probabilities, even when the intention is not just to model completely rational choice voting behaviour \nl{\citep{eggers21b,mebane19,bendor11}}. Importantly, though, this exercise is valuable even for a reader who does not grant that people might vote strategically, so long as they see value in knowing how exactly an IRV ballot could change the results of an election: the pivotal probability derivation requires writing down the set of all pivotal events in IRV, and even if nobody uses this set to vote strategically, it is still an underlying and informative truth about that electoral system.

In that light, this paper makes three advances in our understanding of IRV. The core contribution is to extend the classic calculus of voting to IRV. I derive the probability that a ballot cast in an IRV election will change the outcome of the election, for the first time covering elections with any number of candidates and any length of ballot. This is the central ingredient in any rational choice study of IRV voting. I also show how to estimate pivotal probabilities in IRV using pseudocode, and in the appendix I illustrate and explain the abstract derivation by providing simple numerical examples. After obtaining the full expression for pivotal probability, I pause to consider how complicated it turns out to be. I argue that this is not just a reflection of strategic voting being complicated, but rather that the equation brings into relief something deeply complicated in the idea of instant runoffs.

That derivation also highlights a counter-intuitive property of IRV that I call ``indirect pivotality''. In IRV, one way for a voter to change the outcome of the election is to rank a candidate somewhere on their ballot, and thereby causes that candidate to win the election. However, in IRV, there is another possibility: by ranking some candidate on their ballot, a voter can thereby cause \textit{a different candidate} to win the election. This phenomenon is closely related to some well-studied phenomena (especially IRV's failure of the monotonicity criterion, and also the no-show paradox), but it is not the same as either one. Greater opportunities for indirect pivotality in many-candidate elections may provide a motivation for limiting the number of candidates that voters can rank in IRV elections.

Finally, I compare the probability of casting a pivotal vote in IRV to the probability of casting a pivotal vote in SMDP, and I find that pivotal probabilities in these two systems are very similar. This suggests that voters do not have a larger incentive to behave strategically in one system than the other.

\section{Pivotal probabilities in Instant Runoff Voting}
For about a century IRV was almost uniquely used in Australian legislative elections, but now it is suddenly being rapidly considered and even implemented in large democracies. A 2011 British referendum proposed adopting IRV countrywide, and IRV is reportedly the preferred system of Canada's prime minister \nl{\citep{kohut16}}. The United States is experiencing a surge of IRV adoptions, and this sytem now selects state and federal representatives in Alaska, congressional representatives in Maine, and municipal officials in a number of cities, including the country's largest. One major stated motivation for adopting IRV is that it reduces strategic behaviour, and ``liberates citizens to vote for the candidates they actually favor'' \nl{\citep[ch. 5]{gehl20}}. But attempts to measure strategic voting in IRV do not clearly show that the system reduces or eliminates strategic behaviour, and an even more basic question has gone un-answered: do voters actually have less reason to vote strategically in IRV than in other systems, or are their strategic opportunities similar or even larger?

Single-winner IRV elections work as follows: in an IRV race with $\kappa$ candidates, voters are allowed to rank some number of those candidates. The election administrator counts the number of times that each candidate was ranked first, and the candidate with the fewest first-place votes is eliminated. Then, any remaining candidate that was ranked on a ballot immediately after the candidate that was eliminated has one vote added to their vote total. Again, whichever candidate has the fewest votes is eliminated. This procedure is repeated until $\kappa-1$ candidates have been eliminated, and the remaining candidate wins the election.

In the following section, I will derive the probability that a ballot cast in an IRV election will change the outcome of the election. We will take the perspective of a voter who has only three pieces of information: they know how many candidates are running, they know how many candidates voters are allowed to rank on their ballots, and they know the number of times that they expect the candidates to be ranked in a given order (as if, for example, they had consulted a poll of the whole electorate). I will first establish some notation, and introduce two necessary assumptions. Then, I will derive expressions for how many votes a candidate is expected to receive after a certain sequence of candidates has been eliminated. I will then identify the events in which a voter has pivotal opportunities. Finally, I will model the probability of those events.

\subsection{Notation and assumptions}
First I define some notation. In the following derivations I will denote an ordered slice of a list $\lambda$ from index $a$ up to index $b$ (inclusive) by $\lambda_{a:b}$, indexing from 1, where such a slice can be empty if no elements satisfy the stated requirements. Let $C$ be the set of all $\kappa$ candidates contesting the election. Denote by $v_c$ the expected vote total of candidate $c$, and use $\mu_{a}^{b}|S_{1:n}$ to represent the number of voters expected to rank candidate $a$ in any of the ballot positions between 1 and $b$, given that they assigned every higher ballot position to some candidate in the sequence $S_{1:n}$, which I use to denote the first $n$ elements of the sequence of previously dropped candidates $S$. Also denote the last candidate dropped, $S_{\kappa - 1}$, simply by $S_{-1}$. When considering a sequence of candidates dropped before some candidate $c$ of interest, I will use $A \equiv [S_1, S_2, \cdots S_{\kappa - 1}, c]$. I also denote the ordering of candidates on a ranked-choice ballot by $\beta$, where a voter may make $L$ choices, and the positions are indexed by $i$. The word ``round'' denotes each act of comparing the vote totals of candidates, so that $S_1$ is dropped in the first round, $S_2$ is dropped in the second round, and so on.

I also take the following two assumptions. Substantively, these assumptions are not entirely reasonable; however, I will argue that a) they are at least as reasonable as any alternative assumptions, b) they are necessary for exposition, and c) their limitations do not in any way seriously limit the article's results.

For any candidates $A, B, C, \text{ and } D$ among the $\kappa$ candidates in the election, with $v_a^b$ denoting the vote total of candidate $a$ in round $b$, and using $E_1 \perp E_2 \perp E_3$ to mean that events $E_1, E_2, \text{ and } E_3$ are pairwise independent, \\

\nml \textbf{Assumption 1}: $\big[v_A^r \geq v_B^r \big] \perp \big[v_A^r \geq v_C^r\big] \perp \big[v_C^r \geq v_D^r\big]$

\nml \textbf{Assumption 2}: $\big[v_{A}^r \geq v_{B}^r\big] \perp \big[v_{A}^{r'} \geq v_{B}^{r'}\big] \perp \big[v_{A}^{r'} \geq v_{C}^{r'}\big] \perp \big[v_{C}^{r'} \geq v_{D}^{r'}\big]$ \\

\nml where $A \neq C$, $A \neq D$, $B \neq C$, and $B \neq D$.

Assumption 1 says that any two pairwise comparisons of vote totals are independent, even conditional on any set of candidates having been dropped and any relative ordering of other vote totals. Assumption 2 says that the vote totals of two candidates in some round is unrelated to the vote totals of those or other candidates in another round. Their usefulness to the paper is as follows: Assumption 1 allows us to multiply the probability of two relative vote totals being observed \textit{within} a round and argue that it represents the joint probability of those orderings, while Assumption 2 allows us to obtain a joint probability by multiplying relative vote totals \textit{across} rounds.

Assumption 1 is a feature of the Poisson voting games framework that I will rely on to generate specific probabilities, and has been discussed and used extensively in that context \nl{\citep{myerson98,mebane19,vasselai21b}}. Assumption 2 is more challenging, because it is not a familiar assumption, and it is not obvious why we should accept it. However, I argue that it is at least as reasonable as any alternative, and doing away with it would introduce enormous complication without adding any insight.

Clearly, knowing (for example) the round 1 vote totals of two candidates does give us information about what their round 2 vote totals will be. But what information? To state how likely a candidate is to win the election given that they were leading by a certain margin in round 1 votes, we would need to know the probability that a certain round 1 vote margin will translate into a certain round 2 vote margin, the probability that some round 2 vote margin will turn into some round 3 vote margin, and so on. That probability presumably varies dramatically from election to election, based on arbitrarily complicated features: voters' preferences, campaign effects, coalition dynamics, and so on. Rather than limiting our attention to one such probability, I take the simplest assumption instead: that the vote totals are independent across rounds.

This is \textit{not} a fundamental methodological limitation; it just makes exposition possible. A reader who wishes to do away with either assumption can simply pick their preferred joint probability distribution and substitute it into the derivation in the place of multiplication.\footnote{If a reader wishes to do away with Assumption 1, they cannot obtain the joint probability of two vote totals having certain relative sizes within the same round by simply multiplying the probability that each ordering is realized. Likewise, without Assumption 2, they cannot obtain the joint probability of observing that some vote total is larger than another in round 1, and that also some vote total is larger than another in round 2, by simply multiplying those probabilities. Instead, they must propose exactly what the probability is that the first pair has some relative size \textit{and} the second pair has some relative size. So, wherever the probabilities of relative orderings are multiplied either within rounds or across rounds, they need only introduce a term that gives the probability of these two events occurring. This will produce results identical to the probability that I derive in this paper, up to multiplication of each term by some real number.} However, that would impose an assumption that is no more realistic and at least as strong as Assumption 2, while also severely complicating the derivation.

Two structural points of the derivation should also be mentioned. First, I set aside ties between more than two candidates, which are extremely unlikely in large-population elections and would make the derivations infeasibly more complicated. In SMDP such ties are known to be negligible in large elctorates, except in cases of extremely low turnout \nl{\citep{vasselai21b}}. Second, all probabilities and utilities should be understood as being implicitly compared to abstention. That is, when I discuss (for example) how much more likely a ballot is to be pivotal when candidate $A$ is ranked in position $i$, I am not considering the effects of ranking candidate $A$ \textit{as opposed to some other candidate}, together with the resultant change in that other candidate's vote total --- I only consider the effect of adding one $i$th-place vote to $A$.  

With the notation, assumptions, and structural details in place, we can now derive the pivotal probability of a ballot cast in Instant Runoff Voting.

\subsection{Expected vote totals}
To proceed, we first need an expression for the vote total of a candidate in a round. After $d$ candidates have been dropped, candidate $c$ has the following vote total:

\ml \[v_{c} = \overset{c \ \text{first}}{\underset{\text{vacuous condition}}{\overbrace{\mu_{c}^1}\underbrace{|S_{1:0}}}} + \overbrace{\mu_{c}^{2}|S_{1:1}}^{\substack{c \ \text{second}, \\ \text{any dropped} \\ \text{candidate first}}} + \cdots + \underbrace{\mu_{c}^{d+1}|S_{1:d}}_{\substack{c \ \text{ranked in} \ d + 1, \\ \text{all candidates ranked} \\ 1 \ \text{to} \ d \ \text{dropped}}} \]

\ml \[v_{c} = \sum_{q = 0}^{d} \mu_{c}^{\ d+1} | S_{1:q}\] \ems

\nml A ballot might satisfy more than one of these conditions, but the intention of this equation is that ballots should not be double-counted. Now, denote the probability that candidate $c$ has $k$ more votes than some candidate $j$ by

\ml \[\mathbb{P}(v_c - v_j = k)\] \ems

\nml I will proceed by considering two mutually exclusive ways that a ballot could be pivotal. On the one hand, it could be pivotal by causing a candidate who is ranked on the ballot to win the election. I will call this \textbf{direct pivotality}. On the other hand, placing a candidate in a certain position on a ballot can change the winner of the election without causing that specific candidate to win, which I will call \textbf{indirect pivotality}. The idea of indirect pivotality is closely related to the well-known facts that IRV violates the monotonicity and participation criteria \nl{\citep{lepelley96,ornstein14,smith73}}, but these ideas are distinct for several reasons, and to my knowledge this paper is the first to study indirect pivotality.\footnote{In the social choice literature on runoff systems, pivotal events are often deliberately not studied: it is common to assume for simplicity that ties \textit{do not occur}. See for example \nl{\citet[p. 136]{lepelley96}}.} Let us first consider only direct pivotality, and later return to indirect pivotality.

\subsection{Direct pivotality}
\nml The probability that placing some candidate $c$ in a particular position $i$ on the ballot will cause that candidate to win is the probability that, after every candidate has been eliminated except two, $c$ is among those two remaining candidates, is ranked on the ballot above the other remaining candidate, and is either a) one vote short of winning (breaking a first-place tie), or b) two votes short of winning and would win a tie-breaker (making a first-place tie). The probability that $c$ and some other candidate have the same vote totals after some number of eliminations is:

\ml \[\mathbb{P}(v_c - v_{S_{-1}} = 0)\] \ems

In order to reach a pivotal contest against candidate $S_{-1}$, candidate $c$ must exceed the vote total of every other candidate at the time at which they are dropped. But this is only a pivotal event if $c$ also is not part of the set of dropped candidates $S$. 

By Assumption 2, the probability $p_d|S$ of a directly pivotal contest involving candidate $c$ (for now just considering the case in which a tie can be broken), given a specific sequence $S$ in which the other candidates are dropped, is as follows:

\ml \[p_d|S = \mathbb{P}(v_c = v_{S_{-1}}) \cdot \bigg[\overset{\substack{\text{Probability $c$} \\ \text{not eliminated 1st}}}{\overbrace{\big(\mathbb{P}(\mu_{c}^{1} > \mu_{S_1}^1)\big)}} \underset{\substack{\text{Probability $c$} \\ \text{not eliminated 2nd}}}{\underbrace{\big(\mathbb{P}(\mu_{c}^2|S_1 > \mu_{S_2}^2|S_1)\big)}} \cdots \overset{\substack{\text{Probability $c$ not} \\ \text{eliminated 2nd-last}}}{\overbrace{\big(\mathbb{P}(\mu_c^{\kappa-2}|S_{1:\kappa-3} > \mu_{S_{\kappa-2}}^{\kappa-2}|S_{1:\kappa-3}\big)}} \bigg]\]

\ml \[p_d|S = \mathbb{P}(\mu_c^{\kappa-1}|S_{1:-1} = \mu_{S_{-1}}^{\kappa-1}|S_{1:-1}) \cdot \prod_{h = 1}^{\kappa - 2} \mathbb{P} \big(\mu_c^h|S_{1:h-1} > \mu_{S_h}|S_{1:h-1}\big) \] \ems

\nml This equation is conditional on the candidates being dropped in the order of some list $S$, but the probability that some $S$ is indeed the order in which candidates are dropped is an event that has a probability: namely, the probability that candidate $S_1$ has fewer initial votes than any other candidate, and that $S_2$ has the fewest votes once $S_1$ has been dropped, and so on.

A word must also be said about tie-breakers. An alternative path to the one described above is that $S_1$ could have the same number of votes as $S_2$, and then lose a tie-breaker. I omit this \textit{only} for the sake of exposition, because it is completely infeasible to write out all the comparisons with equality included. Numerical computations of the pivotal probability would have to include every tie-breaking event that could produce a tied vote total between $c$ and $S_1$. Separately, note that casting a vote which breaks a tie between $c$ and $S_1$ is not necessarily pivotal. It is only pivotal in the event that $c$ would have lost the tie-breaker against $S_1$. This tie-breaking probability is feasible to represent, so we include a factor of $\frac{1}{2}$ to represent the event that $c$ would lose a fair tie-breaker without the pivotal vote boosting them into first place.

By Assumptions 1 and 2, the probability $p_d(A)$ that $S$ is the list of candidates dropped, and that at the end of that chain $c$ is involved in a directly pivotal contest, is:

\begin{align*}
	p_d(A) = & \mathbb{P}\big(v_{S_1}^1 < v_{S_2}^{1}\big) \mathbb{P}\big(v_{S_1}^1 < v_{S_3}^{1}\big) \cdots \mathbb{P}\big(v_{S_{1}}^1 < v_{S_{-1}}^{1}\big)\mathbb{P}\big(v_{S_1}^1 < v_{c}^{1}\big) \cdot \\
	& \mathbb{P}\big(v_{S_2}^2 < v_{S_3}^{2}\big) \cdots \mathbb{P}\big(v_{S_{2}}^2 < v_{S_{-1}}^2\big)\mathbb{P}\big(v_{S_2}^2 < v_{c}^2\big) \cdot \\
	& \vdots \\
	& \mathbb{P}\big(v_{S_{\kappa-2}}^{\kappa-2} < v_{S_{-1}}^{\kappa-2}\big)\mathbb{P}\big(v_{S_{\kappa-2}}^{\kappa-2} < v_{c}^{\kappa-2}\big) \cdot \\ 
	& \mathbb{P}\big(v_{S_{-1}}^{\kappa-1} = v_{c}^{\kappa-1}\big) \cdot \frac{1}{2} 
\end{align*}

\ml \[p_d(A) = \bigg[\overset{\substack{\text{Probability the candidates} \\ \text{are dropped in order $A$}}}{\overbrace{\prod_{\ell = 1}^{\kappa - 2} \prod_{r=\ell+1}^{\kappa} \prob \big(v_{A_\ell}^\ell < v_{A_r}^\ell \big)}} \bigg] \overset{\substack{\text{Probability that $c$ is} \\ \text{in a first-place tie} \\ \text{given the drop order $A$}}}{\overbrace{\prob \big(v_{A_{\kappa-1}}^{\kappa-1} = v_c^{\kappa-1}\big)}} \ \overset{\substack{\text{Probability} \\ \text{after tying} \\ \text{$c$ loses}}}{\cdot \overbrace{\frac{1}{2}}} \]

\ml \[p_d(A) = \bigg[\prod_{\ell = 1}^{\kappa - 2} \prod_{r=\ell+1}^{\kappa} \prob \bigg(\sum_{q=0}^{\ell-1} \mu_{A_{\ell}}^{q+1} | A_{1:q} < \sum_{q=0}^{\ell-1} \mu_{A_{r}}^{q+1} | A_{1:q} \bigg) \bigg] \prob \bigg(\sum_{q=0}^{\kappa-2} \mu_{A_{\kappa-1}}^{q+1} | A_{1:q} = \sum_{q=0}^{\kappa-2} \mu_{c}^{q+1} | A_{1:q}\bigg) \cdot \frac{1}{2} \] \ems

The drop sequence $S$ is mutually exclusive with any other sequence, so the probability that ranking candidate $c$ first will be pivotal is

\begin{align*}
	p_d = & \sum_{\substack{S \in \\ \text{Sym}(C \setminus c)}} \bigg\{ \bigg[\prod_{\ell = 1}^{\kappa - 2} \prod_{r=\ell+1}^{\kappa} \prob \bigg(\sum_{q=0}^{\ell-1} \mu_{A_{\ell}}^{q+1} | A_{1:q} < \sum_{q=0}^{\ell-1} \mu_{A_{r}}^{q+1} | A_{1:q} \bigg) \bigg] \cdot \\
	& \frac{1}{2} \cdot \prob \bigg(\sum_{q=0}^{\kappa-2} \mu_{A_{\kappa-1}}^{q+1} | A_{1:q} = \sum_{q=0}^{\kappa-2} \mu_{c}^{q+1} | A_{1:q}\bigg) \bigg\}
\end{align*}

\nml where $\text{Sym}(C \setminus c)$ is the symmetric group of the set of candidates $C$ without $c$.

Finally, one can go beyond the first ballot position by imposing a simple restriction: the $i$th ballot position can only be pivotal if all candidates ranked higher on the ballot have been dropped. 

For the sake of readability I will use $\mu_i$ to mean the vote total of the candidate listed in position $i$ on the length-$L$ ballot $\beta$, and $C \setminus i$ to mean the set of candidates without the candidate listed in position $i$. This yields the following expression for $p_d(\beta)$, the direct pivotal probability of the full ballot $\beta$:

\begin{align*}
	p_{\text{direct}}(\beta) = & \sum_{i=1}^{L} \bigg(\sum_{\substack{S \in \\ \text{Sym}(C \setminus i) \\ \beta_{1:i-1} \subset S}} \bigg\{ \bigg[\prod_{\ell = 1}^{\kappa - 2} \prod_{r=\ell+1}^{\kappa} \prob \bigg(\sum_{q=0}^{\ell-1} \mu_{A_{\ell}}^{q+1} | A_{1:q} < \sum_{q=0}^{\ell-1} \mu_{A_{r}}^{q+1} | A_{1:q} \bigg) \bigg] \cdot \\
	& \frac{1}{2} \cdot \prob \bigg(\sum_{q=0}^{\kappa-2} \mu_{A_{\kappa-1}}^{q+1} | A_{1:q} = \sum_{q=0}^{\kappa-2} \mu_{i}^{q+1} | A_{1:q}\bigg) \bigg\}\bigg)
\end{align*}

Of course, a voter is also pivotal not only if they \textit{break} but also if they \textit{create} a first-place tie, so long as the candidate they elevated into that tie then goes on to win. Note that creating and breaking a tie are mutually exclusive events. However, if a voter creates a first-place tie, there needs to be a tie-breaker. So, now accounting for creating a first-place tie and not just for breaking one, and assuming a fair tie-breaker:

\begin{align*}
	p_{\text{direct}}(\beta) = & \sum_{i=1}^{L} \bigg(\sum_{\substack{S \in \\ \text{Sym}(C \setminus i) \\ \beta_{1:i-1} \subset S}} \bigg\{ \bigg[\prod_{\ell = 1}^{\kappa - 2} \prod_{r=\ell+1}^{\kappa} \prob \bigg(\sum_{q=0}^{\ell-1} \mu_{A_{\ell}}^{q+1} | A_{1:q} < \sum_{q=0}^{\ell-1} \mu_{A_{r}}^{q+1} | A_{1:q} \bigg) \bigg] \cdot \\
	& \bigg[\frac{1}{2} \cdot \prob \bigg(\sum_{q=0}^{\kappa-2} \mu_{A_{\kappa-1}}^{q+1} | A_{1:q} = \sum_{q=0}^{\kappa-2} \mu_{i}^{q+1} | A_{1:q}\bigg) + \\
	& \frac{1}{2} \cdot \prob \bigg(\sum_{q=0}^{\kappa-2} \mu_{A_{\kappa-1}}^{q+1} | A_{1:q} = 1 + \sum_{q=0}^{\kappa-2} \mu_{i}^{q+1} | A_{1:q}\bigg) \bigg] \bigg\}\bigg)
\end{align*}

This is the probability that a ballot cast in an IRV election is directly pivotal for any candidate.

\subsection{Indirect pivotality}
Suppose that candidates will be dropped according to the sequence $A$, but because a voter ranks some candidate $c$ in position $i$ on their ballot, instead candidates are dropped according to a different sequence $A'$. In what situation does the switch from $A$ to $A'$ represent an indirectly pivotal event? 

There are two crucial observations. First, $c$ is not the winner at the end of the sequence $A'$; otherwise, this would be a directly pivotal event (or not pivotal, if $A_{-1} = c$). Second, adding a vote to $c$ cannot change the sequence in which candidates would be dropped \textit{before} $c$.\footnote{This is because I am only considering the pure effects of adding a vote to $c$, as if the elector's only other option were abstention, and not the act of replacing a candidate with $c$. Monotonicity failure, when $A_{-1} = c$ and $A'_{-1} \neq c$, occurs because of this kind of replacement, which increases the vote total of $c$ while simultaneously taking that vote away from another candidate. Also note that indirect pivotality is distinct from a violation of the participation criterion (the ``no-show paradox''), in which a voter can cause a more-preferred candidate to win by abstaining rather than voting. There are two important distinctions: first, indirect pivotality can arise from a voter just changing the order of the ballot they were already going to cast rather than switching from abstention to voting, and second, it does not necessarily cause the victory of a more-preferred candidate.} We can phrase these as constraints on $A$ and $A'$, where $y$ represents the index of $c$ in $A$. \\

\nml \textbf{Constraint 1}: $A$ and $A'$ are the same up to the pivotal event: for some $y$, $A_{1:y-1} = A_{1:y-1}'$ \\

\nml \textbf{Constraint 2}: The winner in $A$ is not the winner in $A'$: $A_{-1} \neq A_{-1}'$ \\

Together these two constraints imply a third: the last element in $A$ cannot be the candidate in $A_{y}$, that is, the winner is some candidate other than $c$. Another simplifying fact is that, for a given pair $A$ and $A'$ (and recalling that I do not consider ties between more than two candidates), whichever candidate $c$ tied with is the candidate that is dropped in that round instead of $c$, so the tie must have been between $c$ and whichever candidate is listed in $A'_y$.

The probability that ranking $c$ in position $y$ will be pivotal is the probability that, for any pair of sequences $A, A'$ satisfying Conditions 1 and 2, the vote for $c$ causes the candidates to be dropped in the sequence specified by $A'$ rather than $A$. Let us seek that probability by first considering just one list $A$.

There must be a tie between $c$ and one of the candidates which remain in the contest by the time that $c$ is dropped. Because I assume that there is a negligible probability of more than two candidates being tied, these events are mutually exclusive by assumption, so the probability of $c$ being involved in a potentially pivotal tie with another candidate is the sum of the probability of $c$ being in a tie or a near-tie with each remaining candidate $t$:

\begin{align*}
	p_{\text{tie}}|A' = \sum_{t \in  A_{y+1:\kappa}} \bigg[\frac{1}{2} \cdot \prob \big(v_c^y = v_t^y\big) + \frac{1}{2} \cdot \prob \big(v_c^y = v_t^y - 1\big) \bigg]
\end{align*}

To obtain the probability of switching from $A$ to $A'$, it is necessary to know not just the probability that there was a tie to create or break, but also that the end of $A'$ will be the specific sequence that follows creating or breaking that tie involving $c$. That is the probability that every candidate $d$ in $A'_{y+1:\kappa}$ defeats every candidate prior to it. For brevity denote $G \equiv A'_{y+1:\kappa}$, that is, $G$ is the alternate ending in the hypothetical pivotal event, and recall that the candidate that $c$ is in a last-place tie with must be $A'_{y}$. For brevity call $t \equiv A'_y$. Then, by Assumption 1, the joint probability is

\begin{align*}
	p_{\text{tie}}(A'|A) = \overset{\substack{\text{Probability that $G$ is the} \\ \text{sequence after a tie with $c$}}}{\overbrace{\prod_{d = 1}^{|G|} \bigg[ \prod_{h = 1}^{d-1} \prob \big( v_{G_{d}} > v_{G_{h}} \big) \bigg]}} \cdot \overset{\substack{\text{Probability that $c$ enters a tie they will win,} \\ \text{or avoids a tie they would not have won}}}{\overbrace{\bigg[\frac{1}{2} \cdot \prob \big(v_c^y = v_t^y\big) + \frac{1}{2} \cdot \prob\big(v_c^y = v_t^y - 1\big) \bigg]}}
\end{align*}

This is the conditional probability of turning some $A$ into a specific sequence $A'$. But there is not just one valid $A'$. Let $\bf{A}$ denote the set of all $A'$ that, for a given $A$, fulfill Conditions 1 and 2, and note that any two lists $A'$ represent mutually exclusive events. Then the probability of any indirectly pivotal event arising from the ranking of $c$ in position $i$ given that the drop sequence would otherwise have followed $A$ is:

\begin{align*}
	p_{\text{tie}}|A = \sum_{A' \in \bf{A}} \bigg\{ \prod_{d = 1}^{|G|} \bigg[ \prod_{h = 1}^{d-1} \prob \big(v_{G_d} > v_{G_{h}} \big) \bigg] \cdot \bigg[\frac{1}{2} \cdot \prob\big(v_c^y = v_t^y\big) + \frac{1}{2} \cdot \prob\big(v_c^y = v_t^y - 1\big) \bigg] \bigg\}
\end{align*}

To obtain the probability of $A'$ actually arising because of a vote for $c$, it is now necessary to know the probability of the sequence $A$ happening. By Assumptions 1 and 2, the probability $p_{\neg \text{d}}(A \to A')$ that a vote for $c$ changes the sequence from $A$ to $A'$ is:

\begin{align*}
	p_{\neg \text{d}}(A \to A') = & \overset{\text{Probability of $A$ occurring}}{\overbrace{\prod_{\ell = 2}^{\kappa} \bigg[ \prod_{h = 1}^{\ell - 1} \prob \big(v_{A_\ell} > v_{A_{h}} \big) \bigg]}} \ \cdot \\
	& \underset{\text{Probability of any $A'$ arising from $A$, with $c$ tied for being dropped at some point}}{\underbrace{\sum_{A' \in \bf{A}} \bigg\{ \prod_{d = 1}^{|G|} \bigg[ \prod_{h = 1}^{d-1} \prob \big(v_{G_d} > v_{G_{h}} \big) \bigg] \cdot \bigg[\frac{1}{2} \cdot \prob \big(v_c^y = v_t^y\big) + \frac{1}{2} \cdot \prob \big(v_c^y = v_t^y = -1\big) \bigg] \bigg\}}}
\end{align*}

What remains is to sum over the possible sequences $A$, and also conduct the calculation for every ballot location. The one important detail in the sum over different values of $A$ is that the only sequences that should be included are those which involve dropping every candidate listed before $i$ on the ballot. The result is the indirect pivotal probability of the ballot in terms of the candidates' vote totals:

\begin{align*}
	p_{\text{indirect}}(\beta) = \sum_{i = 1}^{L} \bigg( \sum_{\substack{A \in \text{Sym}(C) \\ \beta_{1:i} \subset A_{1:y}}} & \bigg[ \prod_{\ell = 2}^{\kappa} \bigg[ \prod_{h = 1}^{\ell - 1} \prob \big(v_{A_\ell} > v_{A_{h}} \big) \bigg] \\
	& \cdot \sum_{A' \in \bf{A}} \bigg\{ \prod_{d = 1}^{|G|} \bigg[ \prod_{h = 1}^{d-1} \prob \big(v_{G_d} > v_{G_{h}} \big) \bigg] \\
	& \cdot \sum_{t \in  F} \bigg[\frac{1}{2} \cdot \prob \big(v_c^y = v_t^y\big) + \frac{1}{2} \cdot \prob \big(v_c^y = v_t^y - 1\big) \bigg] \bigg\} \bigg)
\end{align*}

Finally, one can substitute the equation for candidates' expected vote totals in terms of the current reported ballots in the population. That yields the full expression for the indirect pivotal probability of the ballot $\beta$:

\begin{align*}
	p_{\text{indirect}}(\beta) = \sum_{i = 1}^{L} \bigg( \sum_{\substack{A \in \text{Sym}(C) \\ \beta_{1:i} \subset A_{1:y}}} & \bigg[ \prod_{\ell = 2}^{\kappa} \bigg[ \prod_{h = 1}^{\ell - 1} \prob \bigg(\sum_{q=0}^{h-1} \mu_{A_{\ell}}^{q+1} \big \vert A_{1:q} \ > \sum_{q = 0}^{h-1} \mu_{A_h}^{q+1} \big \vert A_{1:q} \bigg) \bigg] \\
	& \cdot \sum_{A' \in \bf{A}} \bigg\{ \prod_{d = 1}^{|G|} \bigg[ \prod_{h = 1}^{d-1} \prob \bigg(\sum_{q=0}^{y+d} \mu_{G_d}^{q+1} \big \vert A_{1:q} \ >  \sum_{q=0}^{y+d} \mu_{G_h}^{q+1} \big \vert A_{1:q} \bigg) \bigg] \\
	& \cdot \bigg[\frac{1}{2} \cdot \prob \bigg(\sum_{q=0}^{y} \mu_{c}^{q+1} \big \vert A_{1:q} \ = \sum_{q=0}^{y} \mu_{t}^{q+1} \big \vert A_{1:q}\bigg) \\
	& + \frac{1}{2} \cdot \prob \bigg(\sum_{q=0}^{y} \mu_{c}^{q+1} \big \vert A_{1:q} \ = \sum_{q=0}^{y} \mu_{t}^{q+1} \big \vert A_{1:q} - 1 \bigg) \bigg] \bigg\} \bigg] \bigg)
\end{align*}

\subsection{Full pivotal probability}
I have derived both the direct pivotal probability of a given ballot and the indirect pivotal probability of a ballot. Because these are the only two ways to be pivotal, and they are mutually exclusive, the full pivotal probability $p$ is given by $p = p_{\text{direct}} + p_{\text{indirect}}$. At the same time, of interest is not just the probability that the ballot cast will be pivotal, but actually the expected utility of casting a particular ballot. Expected utility is the product of the probability of being pivotal by how much utility would be obtained from the pivotal event: that latter factor in direct pivotality is $u(c) - u(S_{-1})$, while in indirect pivotality it is $u(A'_{-1}) - u(A_{-1})$.

Given a set $C$ of $\kappa$ candidates contesting the election, and with the distribution of intended ballots commonly known, the expected utility of some ballot $\beta$ is:

\begin{align*}
	u(\beta) = & \sum_{i=1}^{L} \bigg(\sum_{\substack{S \in \\ \text{Sym}(C \setminus i) \\ \beta_{1:i-1} \subset S}} \bigg\{ \bigg[\prod_{\ell = 1}^{\kappa - 2} \prod_{r=\ell+1}^{\kappa} \prob \bigg(\sum_{q=0}^{\ell-1} \mu_{A_{\ell}}^{q+1} | A_{1:q} < \sum_{q=0}^{\ell-1} \mu_{A_{r}}^{q+1} | A_{1:q} \bigg) \bigg] \\
	& \bigg[\frac{1}{2} \cdot \prob \bigg(\sum_{q=0}^{\kappa-2} \mu_{A_{\kappa-1}}^{q+1} | A_{1:q} = \sum_{q=0}^{\kappa-2} \mu_{i}^{q+1} | A_{1:q}\bigg) + \\
	& \frac{1}{2} \cdot \prob \bigg(\sum_{q=0}^{\kappa-2} \mu_{A_{\kappa-1}}^{q+1} | A_{1:q} = 1 + \sum_{q=0}^{\kappa-2} \mu_{i}^{q+1} | A_{1:q}\bigg) \bigg] \bigg\}  \cdot \bigg[u(i) - u(S_{-1})\bigg] + \\
	& \sum_{\substack{A \in \text{Sym}(C) \\ \beta_{1:i} \subset A_{1:y}}} \bigg[ \prod_{\ell = 2}^{\kappa} \bigg[ \prod_{h = 1}^{\ell - 1} \prob \bigg(\sum_{q=0}^{h-1} \mu_{A_{\ell}}^{q+1} \big \vert A_{1:q} \ > \sum_{q = 0}^{h-1} \mu_{A_h}^{q+1} \big \vert A_{1:q} \bigg) \bigg] \\
	& \cdot \sum_{A' \in \bf{A}} \bigg\{ \prod_{d = 1}^{|G|} \bigg[ \prod_{h = 1}^{d-1} \prob \bigg(\sum_{q=0}^{y+d} \mu_{G_d}^{q+1} \big \vert A_{1:q} \ >  \sum_{q=0}^{y+d} \mu_{G_h}^{q+1} \big \vert A_{1:q} \bigg) \bigg] \\
	& \cdot \bigg[\frac{1}{2} \cdot \prob \bigg(\sum_{q=0}^{y} \mu_{c}^{q+1} \big \vert A_{1:q} \ = \sum_{q=0}^{y} \mu_{t}^{q+1} \big \vert A_{1:q}\bigg) \\
	& + \frac{1}{2} \cdot \prob \bigg(\sum_{q=0}^{y} \mu_{c}^{q+1} \big \vert A_{1:q} \ = \sum_{q=0}^{y} \mu_{t}^{q+1} \big \vert A_{1:q} - 1 \bigg) \bigg] \bigg\} \bigg] \\
	& \cdot \bigg[u(A'_{-1}) - u(A_{-1})\bigg] \bigg)
\end{align*}

where $\text{Sym}(C)$ is the symmetric group of the set $C$, $L$ is the length of the ballot $\beta$ such that $L \leq \kappa$, $\mu_{a}^{b}|S$ denotes the number of voters expected to rank candidate $a$ in any of the ballot positions 1 through $b$ conditional on assigning any higher ballot position to candidates in the set of previously dropped candidates $S$, $y$ is the index in the list $A$ of the candidate ranked at position $i$, $t$ is the candidate in $A'_y$, and $\bf{A}$ is the set of all ordered lists $A'$ formed by pre-pending $A_{1:y}$ to $G$, for every $G$ in the symmetric group of the set $C \setminus A_{1:y}$.

What can we blame for the complexity of this equation? We have not made any decisions that make the equation more complicated; in fact, we took Assumptions 1 and 2 specifically to simplify the exposition of the equation, even to a point that strains substantive credibility. One suggestion might be that the idea of calculating a pivotal probability is somehow inherently complicated, but notice that we have not yet introduced anything about probabilities \textit{per se}. Crucially, if we just wrote down the set of events in which one ballot changes the result of an IRV election, it would look almost identical to the equation above. Evidently, the complication of this equation can be attrbituted to IRV itself: the process of iterative eliminations is inherently so complicated that simple-sounding questions (``in which cases can my ballot change the election result?'') yield enormously complicated answers.

\subsection{Modeling the probabilities}
There are many ways to model expected vote totals, but I motivated Assumption 1 (and to a lesser extent Assumption 2) as being especially well-supported by one prominent framework: Poisson voting games \nl{\citep{myerson98}}. For that reason I proceed by suggesting how to compute probabilities in that framework, by extending a derivation of pivotal probabilities in single-vote elections by \nl{\citet{mebane19}}. Importantly, however, we have already seen the full pivotal probability equation before ever making the probabilities numerically specific. So, all of the preceding work can be immediately adapted to any other framework for computing probabilities in voting games, such as the Dirichlet beliefs in an iterated polling framework that have been used in previous work on strategic voting in IRV \nl{\citep{eggers21b}}.

Following \nl{\citet{mebane19}}, if the number of voters is drawn from a Poisson distribution, then the number of voters with a preference ordering following each possible preference ordering will also follow a Poisson distribution with known parameter, and their difference follows the Skellam distribution. So one way of computing the expected utility of a ballot is as follows:

\begin{align*}
	u(\beta) = & \sum_{i=1}^{L} \bigg(\sum_{\substack{S \in \\ \text{Sym}(C \setminus i) \\ \beta_{1:i-1} \subset S}} \bigg\{ \bigg[\prod_{\ell = 1}^{\kappa - 2} \prod_{r=\ell+1}^{\kappa} \sum_{w=0}^{\infty} \skel \bigg(w, \sum_{q=0}^{\ell-1} \mu_{A_{r}}^{q+1} | A_{1:q}, \sum_{q=0}^{\ell-1} \mu_{A_{\ell}}^{q+1} | A_{1:q}\bigg) \bigg] \\
	& \bigg[\frac{1}{2} \cdot \skel \bigg(0, \sum_{q=0}^{\kappa-2} \mu_{A_{\kappa-1}}^{q+1} | A_{1:q}, \sum_{q=0}^{\kappa-2} \mu_{i}^{q+1} | A_{1:q}\bigg) + \\
	& \frac{1}{2} \cdot \skel \bigg(1, \sum_{q=0}^{\kappa-2} \mu_{A_{\kappa-1}}^{q+1} | A_{1:q}, \sum_{q=0}^{\kappa-2} \mu_{i}^{q+1} | A_{1:q}\bigg) \bigg] \bigg\}  \cdot \bigg[u(i) - u(S_{-1})\bigg] + \\
	& \sum_{\substack{A \in \text{Sym}(C) \\ \beta_{1:i} \subset A_{1:y}}} \bigg[ \prod_{\ell = 2}^{\kappa} \bigg[ \prod_{h = 1}^{\ell - 1} \sum_{w=0}^{\infty} \skel \bigg(w, \sum_{q=0}^{h-1} \mu_{A_{\ell}}^{q+1} \big \vert A_{1:q}, \ \sum_{q = 0}^{h-1} \mu_{A_h}^{q+1} \big \vert A_{1:q} \bigg) \bigg] \\
	& \cdot \sum_{A' \in \bf{A}} \bigg\{ \prod_{d = 1}^{|G|} \bigg[ \prod_{h = 1}^{d-1} \sum_{w=0}^{\infty} \skel \bigg(w, \sum_{q=0}^{y+d} \mu_{G_d}^{q+1} \big \vert A_{1:q}, \ \sum_{q=0}^{y+d} \mu_{G_h}^{q+1} \big \vert A_{1:q} \bigg) \bigg] \\
	& \cdot \bigg[\frac{1}{2} \cdot \skel \bigg(0, \sum_{q=0}^{y} \mu_{c}^{q+1} \big \vert A_{1:q}, \ \sum_{q=0}^{y} \mu_{t}^{q+1} \big \vert A_{1:q}\bigg) \\
	& + \frac{1}{2} \cdot \skel \bigg(1, \sum_{q=0}^{y} \mu_{c}^{q+1} \big \vert A_{1:q}, \ \sum_{q=0}^{y} \mu_{t}^{q+1} \big \vert A_{1:q} \bigg) \bigg] \bigg\} \bigg] \\
	& \cdot \bigg[u(A'_{-1}) - u(A_{-1})\bigg] \bigg)
\end{align*}

One final caveat to the idea of computing the expected utility of every ballot is that the goal is for voters in a voting game to be able to select the optimal ballot, but nothing guarantees that one ballot will always have the greatest expected utility. In the Appendix I discuss why this could be an extremely common situation in real elections, and I offer suggestions for how a voter might select a most-preferred ballot in the case that multiple different ballots have the same expected utility.

In the Appendix I also apply the pivotal probability equation to a simple example of an IRV election, and I show step-by-step how to use it to compute numerically specific probabilities of both direct and indirect pivotality for a full IRV ballot.

\section{Pseudocode}
Here I provide pseudocode for computing direct and indirect pivotality. I leave counting the votes non-explicit, but each vote total should represent the vote total relevant to the comparison at hand. I also continue to not explicitly represent the probability of ties arising within drop sequences, though these must be computed. The following algorithms are run on the set of all possible ballots, which we have assumed to be $\mathcal{P}_L(\text{Sym}(C))$, that is, the set of all length-$L$ subsets of all of the permutations of the set of candidates $C$.

\FloatBarrier
\begin{algorithm}
\caption{Direct pivotality}
\begin{algorithmic}[1]
\For {voter in voters}
	\For {$\beta$ in allBallots}
		\State ballotDirPivot $\gets 0$
		\For {$i$ in $[1:L]$}
			\For {$S$ in $\text{Sym}(C \setminus \{S_i\})$}
				\State $A \gets S \cup \{i\}$
				\If {$\beta[1:i]$ in $A$}
					\For {each $\ell$ in $A_{1:\kappa-2}$}
						\State candDropProbs $\gets 1$
						\For {$h$ in $A_{\ell:\kappa}$}
							\State chainProb $\gets \sum_{w=1}^{\infty} \mathcal{S} (w, v_\ell, v_h)$
							\State candDropProbs $\gets$ candDropProbs $\cdot$ chainProb
						\EndFor
					\EndFor
					\State breakTieProb $\gets \mathcal{S}(0, v_i, v_{S_{-1}})$
					\State makeTieProb $\gets \mathcal{S}(-1, v_i, v_{S_{-1}})$
					\State pivotProb $\gets$ (candDropProbs)($\frac{1}{2} \cdot$ breakTieProb + $\frac{1}{2} \cdot$ makeTieProb)
					\State ballotDirPivot $\gets$ ballotDirPivot + 	pivotProb
				\EndIf
			\EndFor
		\EndFor
		\State allBallotPivots[$\beta$] $\gets$ ballotDirPivot
	\EndFor
\EndFor
\end{algorithmic}
\end{algorithm}
\FloatBarrier

\FloatBarrier
\begin{algorithm}
\caption{Indirect pivotality}
\begin{algorithmic}[1]
\For {voter in voters}
	\For {$\beta$ in allBallots}
		\State ballotIndirPivot $\gets 0$
		\For {$A$ in $\text{Sym}(C)$}
			\State $\bm{A}$ $\gets$ set of all permutations of $A$ satisfying the two requirements
			\State baseChainProb $\gets$ 1
			\If {$\beta[1:i]$ in $A_{1:y}$}
				\For {each $\ell$ in $A_{2:}$}
					\For {$h$ in $A_{1:\ell-1}$}
						\State baseChainProb $\gets$ baseChainProb $\cdot \sum_{w=1}^{\infty}(w, v_\ell, v_h)$
					\EndFor
					\For {$A'$ in $\bm{A}$}
						\State $G \gets A'_{y:\kappa}$	
						\State altProb $\gets$ 1
						\For {each $d$ in $G$}
							\For {$d$ in $[1:length(G)]$}
								\For {$h$ in $[1:d-1]$}
									\State altProb $\gets$ altProb $\cdot \sum_{w=1}^{\infty}(w, v_{G_d}, v_{G_h})$
								\EndFor
							\EndFor
						\EndFor
						\State $t \gets A'_y$
						\State breakProb $\gets \mathcal{S}(0, v_i, v_t)$
						\State makeProb $\gets \mathcal{S}(0, v_i, v_t - 1)$
						\State pivotProb $\gets$ chainProb $\cdot$ altProb $\cdot$ ($\frac{1}{2} \cdot$ breakProb + $\frac{1}{2} \cdot$ makeProb)
						\State ballotIndirPivot $\gets$ ballotIndirPivot + pivotProb
					\EndFor
				\EndFor
			\EndIf
		\EndFor
	\EndFor
\EndFor
\end{algorithmic}
\end{algorithm}
\FloatBarrier

The pivotal probability of each contest can be multiplied by the utility the voter would obtain from that result, and then the ballot with the largest expected utility selected, with the caveat that a tie-breaking rule might also be necessary.

\section{Simulated pivot probabilities in IRV and SMDP}
I conclude by implementing the IRV pivotal probability algorithm in Python and simulating the magnitude of pivotal probabilities for identical election setups in IRV and SMDP. Now that we can estimate pivotal probabilities in both systems, we can assess the claim that there is more incentive to vote strategically in one system than in the other.\footnote{To facilitate direct comparison, I simulate SMDP pivotal probabilities using the derivation from \nl{\citet{mebane19}} which I extended to cover IRV.}

I will show results from 100 runs of a model, where each run consists of one IRV contest and one SMDP contest with identical parameters, re-run for elections with 3, 4, and 5 candidates. Preferences are drawn in two ways. One is a uniform distribution, where every candidate has similar numbers of people most-preferring them, second-most-preferring them, third-most-preferring them, and so on. The other is a power law distribution, where half of the population holds one preference order, and the other half has a uniformly random preference order. These preference types are opposite ends of a spectrum: the uniform case is an extremely competitive election, in which every candidate has similar levels of support, while the power law case is a model of an election where there is a clear front-runner. In both IRV and SMDP, pivotal probabilities should be larger in competitive elections than in uncompetitive elections, but it is not obvious in advance how the different preference distributions should increase or decrease pivotal probabilities more in one system than the other.

\begin{figure}[h!]
	\begin{center}
		\subfigure[Pivotal probabilities with power law support]{
			\label{fig:pivot_power}
			\includegraphics[width=0.66\textwidth]{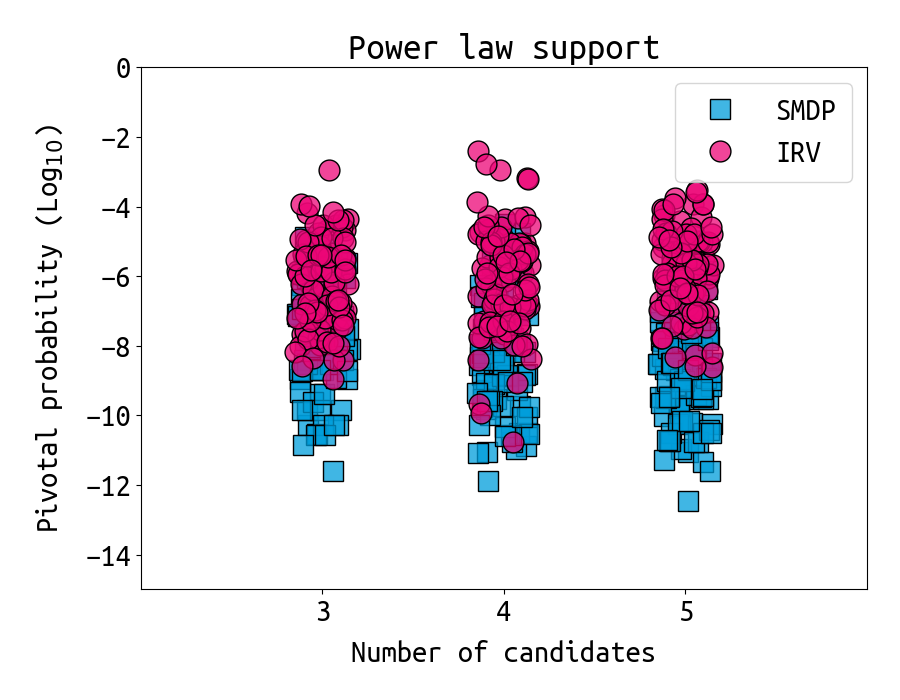}
		}
		\subfigure[Pivotal probabilities with uniform support]{
			\label{fig:pivot_unif}
			\includegraphics[width=0.66\textwidth]{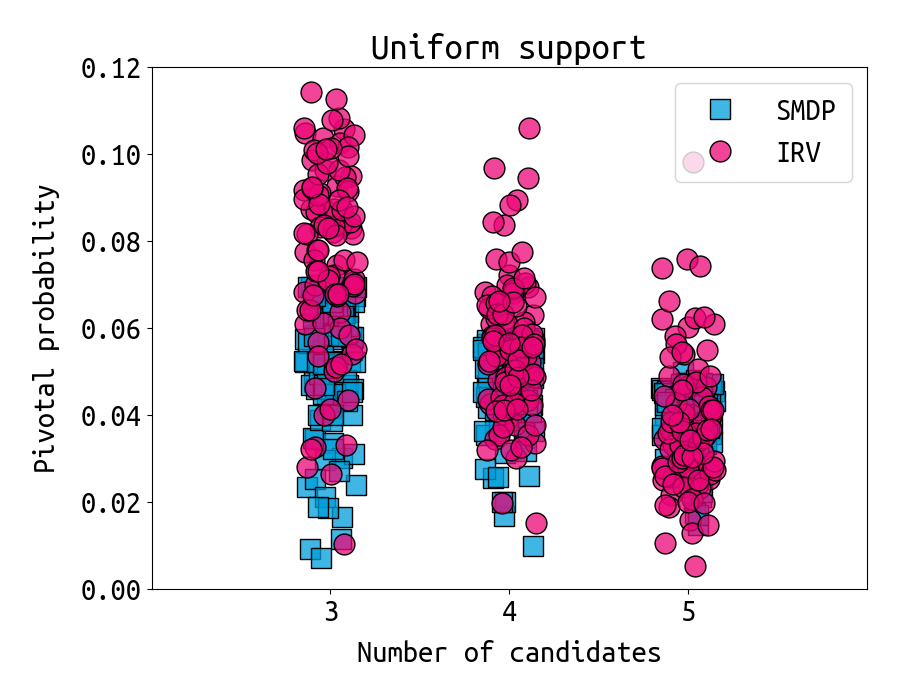}
		} 
	\end{center}
	\caption{\label{fig:pivots} The comparative pivotal probabilities of IRV and SMDP with 3, 4, and 5 candidates, when voters can rank every candidate, in a community of 1,000 voters. The number is obtained by summing the pivotal probability of \textit{all} ballots. Candidates have an equal probability of appearing in any position on any ballot. 100 runs are shown with different random number seeds, and for each run one contest is held under IRV and another contest with identical starting conditions is held under SMDP for each number of candidates.}
\end{figure}

Figure \ref{fig:pivots} shows the total pivotal probability for a given preference distribution in either IRV and SMDP. The total pivotal probability on the $y$-axis is the sum of the pivotal probability of every ballot.

\FloatBarrier
\FloatBarrier

The striking feature of Figure \ref{fig:pivots} is that IRV and SMDP have extremely similar pivotal probabilities. When support for candidates resembles a power law distribution, IRV has very slightly higher pivotal probabilities than SMDP (averaging about $10^{-6}$ compared to about $10^{-7}$ with these particular preference structures). In the case of uniformly distributed support, the chances to be pivotal in IRV may be slightly higher than in SMDP when there are only 3 candidates, but when there are 4 or 5 candidates neither is clearly larger than the other.\footnote{However, I strongly caution against reading substantively into the appearance that pivotal probabilities in IRV fall more quickly as the number of candidates rises than pivotal probabilities in SMDP, since this result may be attributable to the varying quality of Assumption 2 as more candidates are introduced. Also, note that these are all out-of-equilibrium pivotal probabilities. Observing these pivotal probabilities would cause rational voters to update their expected vote choice, which should change the expected vote totals of each candidate, and in turn alter the pivotal probabilities. This process may or may not converge to a fixed equilibrium. 
So, these should be understood as the \textit{initial} pivotal probabilities of these election structures.} The results do not support widespread claims that there are either stronger or weaker incentives for voters to cast strategic votes under IRV compared to SMDP.

\section{Conclusion}
This article has extended the classic calculus of voting to IRV, by deriving the probability that a ballot cast in an IRV election will determine the election outcome. I have shown how to implement that pivotal probability calculation using both pseudocode and examples. This core theoretical contribution led to two substantive results.

First, the notion of pivotal voting has to be extended in an counter-intuitive way to fit IRV. In classic pivotal voting, a voter elevates a candidate into first place by casting a vote for them. In IRV, a voter can rank candidate $A$ somewhere on their ballot, and thereby cause $C$ to win instead of $B$. Indirect pivotality may be rare, but it is easy to provide examples of small-electorate contests in which the election may be decided by an indirectly pivotal vote choice. If hundreds of IRV elections are held in electorates with dozens of voters, it would be reasonable to expect that there will regularly be some voter who caused one candidate to win because they chose to vote for an entirely different candidate. 

Second, by putting numbers to the probability of causing a pivotal event in IRV, I was able to evaluate the widespread expectations that, compared to classic SMDP voting, IRV either provides more opportunities for strategic voting, or else contains lower incentives for strategic voting. I showed that neither of these stories is clearly correct. The probability that some ballot in an IRV election will be pivotal is very similar to the probability that some ballot in an SMDP election will be pivotal. This suggests that the opportunities and incentives for strategic behaviour in IRV are about the same as they are in SMDP.

\newpage

\bibliography{irv}

\begin{thebibliography}{}

\bibitem[Atsusaka, 2023]{atsusaka23}
Atsusaka, Y. (2023).
\newblock Causal inference with ranking data: Application to blame attribution
  in police violence and ballot order effects in ranked-choice voting.
\newblock {\em Working paper}.

\bibitem[Baltz, 2022]{baltz22}
Baltz, S. (2022).
\newblock Simulating electoral system changes and how voters might respond.
\newblock {\em Working paper}.

\bibitem[Bendor et~al., 2011]{bendor11}
Bendor, J., Diermeier, D., Siegel, D.~A., and Ting, M.~M. (2011).
\newblock {\em A behavioral theory of elections}.
\newblock Princeton University Press.

\bibitem[Bouton, 2013]{bouton13}
Bouton, L. (2013).
\newblock A theory of strategic voting in runoff elections.
\newblock {\em American Economic Review}, 103(4):1248--1288.

\bibitem[Buisseret and Prato, 2022]{buisseret22}
Buisseret, P. and Prato, C. (2022).
\newblock Politics transformed? how ranked choice voting shapes electoral
  strategies.
\newblock {\em Working paper}.

\bibitem[Cox, 1994]{cox94}
Cox, G.~W. (1994).
\newblock Strategic voting equilibria under the single nontransferable vote.
\newblock {\em American Political Science Review}, 88:608--621.

\bibitem[Cox and Shugart, 1996]{cox96}
Cox, G.~W. and Shugart, M.~S. (1996).
\newblock Strategic voting under proportional representation.
\newblock {\em Journal of Law, Economics, \& Organization}, 12(2):299--324.

\bibitem[Eggers and Nowacki, 2021]{eggers21b}
Eggers, A.~C. and Nowacki, T. (2021).
\newblock Susceptibility to strategic voting: A comparison of plurality and
  instant-runoff elections.

\bibitem[Eggers and Vivyan, 2020]{eggers20}
Eggers, A.~C. and Vivyan, N. (2020).
\newblock Who votes more strategically?
\newblock {\em American Political Science Review}, 114(2):470--485.

\bibitem[Gehl and Porter, 2020]{gehl20}
Gehl, K.~M. and Porter, M.~E. (2020).
\newblock {\em The politics industry: How political innovation can break
  partisan gridlock and save our democracy}.
\newblock Harvard Business Review Press.

\bibitem[Kohut, 2016]{kohut16}
Kohut, T. (2016).
\newblock What trudeau said: A look back at liberal promises on electoral
  reform.
\newblock {\em Global News}.

\bibitem[Lepelley et~al., 1996]{lepelley96}
Lepelley, D., Chantreuil, F., and Berg, S. (1996).
\newblock The likelihood of monotonicity paradoxes in run-off elections.
\newblock {\em Mathematical Social Sciences}, 31:133--146.

\bibitem[Mebane et~al., 2019]{mebane19}
Mebane, Jr., W.~R., Baltz, S., and Vasselai, F. (2019).
\newblock Using agent-based models to simulate strategic behavior in elections.
\newblock {\em Presented at the 2019 meeting of the Society for Political
  Methodology}.

\bibitem[Myerson, 1998]{myerson98}
Myerson, R.~B. (1998).
\newblock Population uncertainty and poisson games.
\newblock {\em International Journal of Game Theory}, 27:385--392.

\bibitem[Ornstein and Norman, 2014]{ornstein14}
Ornstein, J.~T. and Norman, R.~Z. (2014).
\newblock Frequency of monotonicity failure under instant runoff voting:
  estimates based on a spatial model of elections.
\newblock {\em Public Choice}, 161(1/2):1--9.

\bibitem[Reilly, 2021]{reilly21}
Reilly, B. (2021).
\newblock Ranked choice voting in australia and america: Do voters follow party
  cues?
\newblock {\em Politics and Governance}, 9(2):271--279.

\bibitem[Riker and Ordeshook, 1968]{riker68}
Riker, W.~H. and Ordeshook, P. (1968).
\newblock A theory of the calculus of voting.
\newblock {\em American Political Science Review}, 62(1):25--42.

\bibitem[Santucci, 2021]{santucci21}
Santucci, J. (2021).
\newblock Variants of ranked-choice voting from a strategic perspective.
\newblock {\em Politics and Governance}, 9(2):344--353.

\bibitem[Simmons et~al., 2022]{simmons22}
Simmons, A.~J., Gutierrez, M., and Transue, J.~E. (2022).
\newblock Ranked-choice voting and the potential for improved electoral
  performance of third-party candidates in america.
\newblock {\em American Politics Research}, 50(3).

\bibitem[Smith, 1973]{smith73}
Smith, J.~H. (1973).
\newblock Aggregation of preferences with variable electorate.
\newblock {\em Econometrica}, 41(6):1027--1041.

\bibitem[Vasselai, 2022]{vasselai21b}
Vasselai, F. (2022).
\newblock Pivotal probabilities in iterative voting under plurality.
\newblock {\em AAAI Conference on Artificial Intelligence}.

\end{thebibliography}
\bibliographystyle{apalike}

\end{document}